\documentclass[12pt]{article}
\usepackage{amssymb}
\usepackage{graphicx}
\usepackage{epsfig}
\usepackage{wrapfig}
\newcommand{\lapproxeq}{\lower .7ex\hbox{$\;\stackrel{\textstyle  
<}{\sim}\;$}} 
\newcommand{\gapproxeq}{\lower .7ex\hbox{$\;\stackrel{\textstyle  
>}{\sim}\;$}} 
\newcommand{\stackdown}[2]{\lower 1.4ex\hbox{$\;\stackrel{\textstyle{#1}}  
{\scriptstyle{#2}}\;$}}
\newcommand{\beq}{\begin{equation}} 
\newcommand{\eeq}{\end{equation}} 
\newcommand{\bea}{\begin{eqnarray}} 
\newcommand{\eea}{\end{eqnarray}}

\def\beq{\begin{equation}}
\def\eeq{\end{equation}}
\newcommand{\form}[1]{(\ref{#1})}

\makeatletter 
\def\slash{\@ifnextchar[{\fmsl@sh}{\fmsl@sh[0mu]}} 
\def\fmsl@sh[#1]#2{%
  \mathchoice 
    {\@fmsl@sh\displaystyle{#1}{#2}}%
    {\@fmsl@sh\textstyle{#1}{#2}}%
    {\@fmsl@sh\scriptstyle{#1}{#2}}%
    {\@fmsl@sh\scriptscriptstyle{#1}{#2}}} 
\def\@fmsl@sh#1#2#3{\m@th\ooalign{$\hfil#1\mkern#2/\hfil$\crcr$#1#3$}} 
\makeatother 

\def\s{{\,\rm s}}

\def\beq{\begin{equation}}
\def\eeq{\end{equation}}

\catcode`\@=11 

\def\lsim{\mathrel{\mathpalette\@versim<}}
\def\gsim{\mathrel{\mathpalette\@versim>}}
\def\@versim#1#2{\vcenter{\offinterlineskip
    \ialign{$\m@th#1\hfil##\hfil$\crcr#2\crcr\sim\crcr } }}

\def\t1{{\tilde 1}}

\def\slash#1{#1\hskip-6pt/\hskip6pt}

\def\to{\rightarrow}

\begin{document}

\begin{titlepage}

\begin{flushright}
astro-ph/0308403 \\
CERN-TH/2003-197 \\
FTUV-030822
\end{flushright} 

\vspace{0.5cm}

\begin{centering} 

{\large {\bf Synchrotron Radiation from the Crab Nebula Discriminates 
between Models of Space-Time Foam }}

\vspace{1cm} 

{\bf J.~Ellis$^a$}, {\bf N.E.~Mavromatos$^{b,c}$} and  
{\bf A.S.~Sakharov$^{a,d}$}

\vspace{0.5cm} 

$^a$ {\it CERN, Theory Division, CH-1211 Geneva 23, Switzerland.}

$^b$ {\it Department of Physics, King's College London, University of London, 
Strand , London WC2R 2LS, U.K.} 

$^c$ {\it Departamento de F\'isica T\'eorica, Universidad de Valencia,
E-46100, Burjassot, Valencia, Spain.} 

$^d$  {\it Swiss Institute of Technology, ETH-Z\"urich, 8093 Z\"urich,
Switzerland.}

\vspace{1cm}

{\bf Abstract}

\vspace{0.5cm}

\end{centering} 

It has been argued by Jacobson, Liberati and Mattingly that synchrotron
radiation from the Crab Nebula imposes a stringent constraint on any
modification of the dispersion relations of the {\it electron} that might
be induced by quantum gravity. We supplement their analysis by deriving
the spectrum of synchrotron radiation from the coupling of an
electrically-charged particle to an external magnetic fields in the
presence of quantum-gravity effects of the general form
$(E/M_{QG})^\alpha$. We find that the synchrotron constraint from the Crab
Nebula practically excludes $\alpha  \lsim  1.74$ 
for $M_{QG} \sim m_P = 1.2 \times
10^{19}$~GeV. On the other hand, this analysis does not constrain any
modification of the dispersion relation of the {\it photon} that might be
induced by quantum gravity. We point out that such quantum-gravity effects
need not obey the equivalence principle, a point exemplified by the
Liouville-string D-particle model of space-time foam. This model suggests
a linear modification of the dispersion relation for the photon, but {\it
not} for the electron, and hence is compatible with known constraints from
the Crab Nebula and elsewhere.

\vfill
\leftline{CERN-TH/2003-197}
\leftline{August 2003}
\end{titlepage}
\baselineskip=18pt


\section{Introduction}

Modified dispersion relations have been first suggested in the context of
stringy quantum gravity (QG) in~\cite{aemn}, based
on the Liouville string approach to quantum 
space time~\cite{emnerice}. Later, analogous
modifications have also been suggested in the contexts of other models of
QG, either phenomenological~\cite{mestres} or theoretical. Examples of the
latter include loop gravity~\cite{smolin} and novel models in which the
Planck scale is viewed as a length that remains invariant under modified
non-linear Lorentz transformations~\cite{amelino1}. Violations of Lorentz
symmetry had been suggested earlier~\cite{coleman1} as a way of avoiding
the Greisen-Zatsepin-Kuzmin (GZK) cutoff for ultra-high-energy cosmic rays
(UHECRs).

Liouville-string models of space-time foam~\cite{emnerice} 
motivate corrections to the
usual relativistic dispersion relations that are of first order in the
particle energies, corresponding to a vacuum refractive index $\eta \simeq
1 - (E/M_{QG})^\alpha$: $\alpha = 1$. These effects are associated
generically with deviations from conformal invariance in the effective
theory of low-energy excitations interacting with singular or
topologically non-trivial quantum-gravitational (QG) degrees of freedom,
inaccessible to low-energy observers~\cite{emnerice}. Models with 
quadratic dependences of the vacuum refractive index on energy: $\alpha 
=2$ have also been considered~\cite{burges}.

The phenomenology of such models has grown rapidly.  Following the
original suggestion~\cite{nature,gray} to place bounds on the effective QG
scale by comparing the arrival times of photons with different energies
from gamma-ray bursters (GRBs), in order to probe the refractive index 
that
might be induced by QG, it has also been pointed out that
electrically-charged fermionic probes, either in astrophysics~\cite{piran}
or in terrestrial atomic and nuclear physics
experiments~\cite{others,jlm,stecker}, can constrain severely
phenomenological models of space-time foam with modified dispersion
relations.

The most severe of all the known constraints seems to be that associated
with synchrotron radiation from the Crab Nebula, whose sensitivity exceeds
the Planck scale by (at least) seven orders of magnitude~\cite{jlm}, in
the case of a linear modification of the dispersion relation for the {\it
electron}. For {\it photons}, on the other hand, the most precise probe of
a possible QG-induced refractive index comes from an analysis of arrival
times of emissions from GRBs~\cite{gray}, which impose $M_{QG} \gsim
10^{16}$ GeV. Analyses of light from active galactic nuclei (AGNs) and
pulsars~\cite{kaaretbiller} also reveal no evidence of QG effects, and may 
have comparable
sensitivities to $M_{QG}$ for photons.

In the first part of this paper, we revisit the analysis of~\cite{jlm}, by
considering in detail the propagation of a charged matter probe
interacting with an external magnetic field in the case of a generic
modified dispersion relation, not necessarily linearly suppressed by the
Planck scale. Our analytical result includes explicitly corrections due to
the calculable change in the synchrotron radius in the presence of such QG
effects. We confirm that the synchrotron constraint provided by the Crab
Nebula~\cite{jlm} on the electron's dispersion relation is robust
theoretically. Unless it is relaxed for astrophysical reasons, this type 
of constraint is so strong that it already excludes an $\alpha\le 1.7$
correction to the vacuum refractive index for electrons, and has the
potential to be sensitive to a quadratic correction:  $\alpha = 2$ in the
near future, when higher-precision data become available.

However, this does not mean that all types of QG corrections with $\alpha 
\le 2$ are
excluded. The escape route is for QG to violate the equivalence principle, 
so that $\alpha = 1$ and
$\eta < 1$ for photons, whilst $\eta = 1$ for electrons. Remarkably,
this is exactly what happens in the Liouville model of space-time foam
proposed in~\cite{emn}, in the modern framework of
the D-brane approach to QG~\cite{polchinski}. 
The reason for this violation of the equivalence
principle for different categories of particles is explained briefly at
the end of this paper, and is described in more detail
elsewhere~\cite{EMS2}.

The structure of the article is as follows: in the next Section we
summarize the Crab Nebula analysis of~\cite{jlm}. Then, in Section 3 we
present an analytical derivation of the synchrotron radiation spectrum in
the presence of modified dispersion relations, including the change in the
synchrotron radius that we show to be a small effect. We also show that a
quadratic modification of the electron's dispersion relation with $M_{QG}
= m_P$ is marginally excluded by the Crab data. Finally, in Section 4, we
discuss the ability of this constraint to exclude some models of QG,
showing how the specific model of space-time foam in~\cite{emn} evades it
by violating the equivalence principle in a characteristic way.

\section{The Synchrotron Radiation Constraint from the Crab Nebula}

Let us briefly summarize the main points of~\cite{jlm}, who first proposed
considering the constraints implied by synchrotron radiation from the Crab
Nebula. We highlight some steps which were not made explicit in their 
analysis,
motivating our derivation in the next Section of an analytical
mathematical description of synchrotron radiation in the presence of
modified dispersion relations.

Following~\cite{jlm}, we
assume the modified dispersion relations (in units of the 
speed of light in vacuo $c$, which we now set to unity)
\begin{eqnarray}
 \omega^2(k)&=& k^2 + \xi_\gamma \frac{k^3}{M_P},
 \label{eq:pdr}\\
 E^2(p)&=& m_0^2+ p^2 + \xi_e \frac{p^3}{M_P},
 \label{eq:mdr}
\end{eqnarray}
for photons and electrons, respectively, where $\omega$ and $k$ are the
photon frequency and wave number, and $E$ and $p$ are the electron energy
and momentum, with $m_0$ the electron rest mass. Here we assume linear QG
effects, characterized by parameters $\xi_\gamma$ and $\xi_e$, extracting
the Planck mass scale $m_{P}=1.22\times 10^{19}$ GeV. We also assume
here that energy and momentum are conserved in particle
interactions~\footnote{This may be questioned~\cite{ng,emngzk}, 
but any plausible violation
is unimportant for our purposes.}. We expect that the parameters $\xi_\gamma$
and $\xi_e $ are negative semi-definite, given that in the framework of
Liouville-inspired quantum gravity~\cite{emn} there are at most {\it
subluminal} modifications, if any, and superluminal modifications would
cause troubles with gravitational {\v C}erenkov radiation~\cite{nelson}. 
We note that
the modifications to the dispersion relation proposed in~\cite{emn} arise
from a {\it non-Minkowski} induced metric in target space, which leads to
some formal differences from the approach of \cite{jlm}, as we comment
later. 

We now recapitulate the derivation of constraints from the
Crab Nebula
synchrotron radiation on phenomenological models of quantum gravity that
incorporate (\ref{eq:mdr}) for electrons.  It was assumed in~\cite{jlm}
that the usual description of synchrotron radiation was applicable, and
the following facts about spectrum of the Crab nebula were used:
\begin{itemize}
\item{}  Synchrotron emission is observed up to energies of 
about $0.5$ GeV, where the inverse Compton hump begins to dominate the 
spectrum.

\item{} Photons with energies up to 50
TeV~ are observed.

\item{} Electrons with energies
up to at least 50 TeV are required by energy conservation to produce
the observed flux of 50 TeV photons by inverse Compton scattering.
\end{itemize}
In the standard Lorentz-invariant theory, the 0.5 GeV synchrotron
emission in the magnetic field of the Crab nebula ($\sim 0.3$
mG) is generated by electrons of energy $5\times
10^4$ TeV.  The inference of this energy assumes, however, Lorentz
invariance, which is just what we want to test.  Hence the 
authors of~\cite{jlm} adopted the more conservative
lower value of 50 TeV, which is inferred using only energy
conservation in the inverse Compton process.

In standard electrodynamics, accelerated electrons in a magnetic field
emit synchrotron radiation with a spectrum that cuts off sharply at a
frequency $\omega_c$ which can be calculated as follows. 
According to the standard theory, electrons in an
external magnetic field $H$, follow helical orbits
transverse to the direction of $H$, 
with the orbit radius $R$ given by
\begin{equation} 
 R = \frac{E}{H}
\label{radius}
\end{equation}
where $E$ is the total energy of the electron. 
The orbital frequency of the electron in this case is:
\begin{equation} 
\omega _0 = \frac{\beta_\perp}{R}
\label{freqmf}
\end{equation}
where $\beta_\perp \equiv v_\perp $ is the component of the 
velocity of the electron perpendicular to the direction of the 
magnetic field. It was argued in~\cite{amelino}
that there could be modifications to $R$ that might affect the results
of~\cite{jlm}. We show later that there are 
indeed QG induced modifications, 
but that these 
do not change the central result of~\cite{jlm}.

We recall that, in the standard Lorentz-invariant (LI) 
theory of synchrotron radiation, an 
electron moving in a magnetic field $H$ emits a discrete 
spectrum of  frequencies, which are integer multiples of:
\beq
\label{multiple}
\omega_0^*=\frac{\omega_0}{sin^2\theta},
\eeq
where $\theta$ is the angle between the velocity of the electron, $\vec 
v$, and the direction of $\vec H$. The spectrum of the emitted radiation 
has a maximum at a critical frequency 
\begin{equation} 
\omega^{LI}_c = \frac{3}{2}\frac{eH}{m_0}\frac{1}{1 - \beta^2},
\label{sync}
\end{equation}
where $e$ is the electron charge. The superfix $LI$ in (\ref{sync})
stresses that this formula is based on a LI approach, in which one 
calculates the electron trajectory in a given magnetic field $H$ and the
radiation produced by a given current, using the relativistic relation
between energy and velocity. All of these assumptions could in principle 
be affected by violations of Lorentz symmetry.

It was stated in~\cite{jlm} that, even
without assuming Lorentz invariance, the critical frequency is given by
\begin{equation}
\omega_c=\frac{3}{4} \frac{1}{R \delta(E)}\;
\frac{1}{c(\omega_c)-v(E)} \label{eq:opeaktilde}
\end{equation}
where $\delta(E)$ is the opening angle for the forward-directed radiation
pattern, and $c(\omega_c)$ and $v(E)$ are the group velocities of the
radiation and electron respectively. The self-consistent solution of
(\ref{eq:opeaktilde}) for $\omega_c(E)$ determines the cutoff synchrotron
frequency. There was no attempt to solve this equation in~\cite{jlm}.
Instead, it was argued that one can replace $c(\omega_c)$ by $c = 1$, 
so that the
self-consistent solution is simply equal to the right-hand side of
(\ref{eq:opeaktilde}). To motivate this result, one may note that the
electron and photon speeds are very close to the low-energy speed of light
in vacuo $c=1$.

The synchrotron radius $R$ for any given energy is determined by the 
equation of
motion of the electron. The authors of~\cite{jlm} 
assumed that gauge invariance is preserved, and used the usual minimal 
coupling. To find the electron
equation of motion in a magnetic field ${H}$, they used 
the dispersion relation (\ref{eq:mdr})
for the Hamiltonian, with the momentum replaced by 
\begin{equation} 
{\bf p} \to {\bf p}  - e{\bf A}
\label{covariant}
\end{equation}
where ${\bf A}$ is a vector potential for the magnetic field. This yields
the equation of motion ${\bf a}=[1 + 3\xi_e E/2M](e/E)\, {\bf v}\times{\bf
H}$, where only the term of lowest order in $\xi_e$ is kept, when one
assumes $E\gg m_0$~\footnote{We note that the covariantization
(\ref{covariant}) was done assuming that the background space-time is flat
Minkowski space, as in the phenomenological analysis of \cite{jlm}.}.
Since $E\ll M$, the authors of \cite{jlm} argued that Lorentz violation
would make very little difference to the orbital equation, so that the
radius $R$ would be related to the magnetic field $H$ and the energy $E$
of the electron by the standard formula (\ref{radius}).
These arguments have been refuted 
qualitatively in~\cite{amelino}, with a subsequent reply 
in~\cite{repljac}. 
In order to arrive at a decisive conclusion, we embark in this 
article on a detailed quantitative 
analysis of the QG modification of the curvature radius.

In the next Section, we revisit the analysis of~\cite{jlm}, presenting a
derivation that includes the modification of this formula for the orbit
radius due to the modification of the dispersion relation. The potential
importance of this modification has been stressed in~\cite{amelino}, but
we find it to be relatively small. Our calculations should be accurate for
all energies $E \lsim 0.3~M_P/|\xi_e|$.
The analysis of~\cite{jlm} can then be used to infer a strong
lower bound of the effective quantum gravity scale {\it for electrons}:
$M_P/|\xi_e| \sim 10^{27}$ GeV.  However, we reiterate that there are no
such strict bounds for photons, for which the most stringent bounds on the
quantum gravity scale $M_P/|\xi_\gamma| \gsim 10^{16}$ GeV have been 
derived
from gamma-ray bursts~\cite{gray}.

One might be tempted to assume universality for the QG corrections
$\xi_\gamma$ and $\xi_e$, as a consequence of the equivalence principle,
in which case the electron bound would exclude observable linear QG
modifications of dispersion relations in general. However, one should not
exclude {\it a priori} the possibility that the equivalence principle
might be violated for such QG corrections, with different effective
quantum scales for different particle species, as considered
in~\cite{coleman1,glashow}.  We argue in Section 5 that this is indeed the
case in the Liouville model for space-time foam of~\cite{emn}, where
electric charge conservation makes the foam medium transparent to charged
probes such as electrons, but not to photons.

Before proceeding with our detailed derivation of the modified synchrotron
spectrum in the next Section, we first recall how (3) was used 
in~\cite{jlm} to derive synchrotron radiation bounds.
Using the
dispersion relations (\ref{eq:pdr},\ref{eq:mdr}), and keeping terms
suppressed only linearly by the QG scale $M_P$, one may write
the difference of group velocities in the denominator of the last term of
(\ref{eq:opeaktilde}) as:
\begin{equation}
 c(\omega)-v(E)= \xi_\gamma \omega + (m_0^2/2E^2) - \xi_e E,
 \label{eq:vdiff}
\end{equation}
The synchrotron radiation constraint comes from maximizing the 
electron group velocity with respect to the energy $E$, which yields
for the maximum photon frequency
~\cite{jlm}:
\begin{equation}
\omega_c=\frac{3}{2} \frac{eH}{m_0}\frac{m_0\gamma(E)}{E}\;
\gamma^2(E). \label{eq:opeaklv}
\end{equation}
Comparing (\ref{eq:opeaklv}) with (\ref{sync}), it was observed
in~\cite{jlm} that the factor $m_0\gamma(E)/E$, which is different from
unity as a consequence of Lorentz violation, is a bounded function of $E$,
\begin{equation}
 \gamma(E)=(1-v^2)^{-1/2}\approx
  \left(\frac{m_0^2}{E^2} - 2\xi_e \frac{E}{M} \right)^{-1/2}.
   \label{eq:gm}
\end{equation}
Maximizing $\omega_c$ with respect to the energy $E$ yields
\begin{equation}
\omega_c^{\rm max}=0.34 \, \frac{eH}{m_0}(-\xi_e
m_0/M)^{-2/3},
\label{eq:opeaklv2}
\end{equation}
which is attained at the energy $E_{\rm max}=(-2m_0^2M/5\xi _e)^{1/3}=10\,
(-\xi _e)^{-1/3}$ TeV.  The frequency $\omega_c^{\rm max}$ is the highest
possible value of the critical synchrotron frequency for any electron
energy.  The rapid decay of synchrotron emission at frequencies larger
than $\omega_c$ implies that most of the flux at a given frequency in the
synchrotron radiation peak observed coming from the Crab Nebula is due to
electrons for which $\omega_c$ is above that frequency. Thus
$\omega_c^{\rm max}$ must be greater than the maximum observed synchrotron
emission frequency {\bf $\hbar \omega_{\rm obs}$}, which
yields the constraint
\begin{equation}
\xi _e >  -\frac{M}{m_0}\left(\frac{0.34\, eH}{m_0\omega_{\rm
obs}}\right)^{3/2}. \label{eq:synchcon}
\end{equation}
This was the main result of~\cite{jlm}. It implies, when one takes into
account indicative physical values/estimates of the various quantities
entering (\ref{eq:synchcon}), a very small upper bound on the magnitude of the
coefficient $|\xi_e|$ of the QG modification of the electron dispersion
relation (which is assumed to be negative, corresponding to subluminal
propagation):
{\bf \begin{equation}
|\xi _e|  < 7\times 10^{-8}.
\label{eq:numba2}
\end{equation}}
As already mentioned, this analysis does not yield a stringent constraint
on the magnitude of the photon coefficient $|\xi _\gamma|$.

At this stage, we remark that, in the framework of our Liouville model of
space-time foam, we have derived QG modifications only for massless
neutral particles such as the photon and photino, the latter being used as
a model for the neutrino~\cite{EMNV}. As discussed later in this paper, 
different
categories of particles may have different QG modifications in their
dispersion relations, and these derivations do not apply to electrons.

\section{Analytic Derivation of the Synchrotron Radiation Constraint}

We now present an analytical derivation of the QG modification to 
the synchrotron-radiation formula, for generic subluminal 
modifications of the electron dispersion relation of the form:  
\begin{equation} 
 E^2(p) = m_0^2+ p^2 - \frac{p^{2+\alpha}}{{\mathcal M^\alpha}}~,
\label{higher} 
\end{equation}
where ${\cal M}$(=$M_P/|\xi _e|$ for $\alpha =1$)  
sets the effective QG scale in the
modifications of the dispersion relation for the electron, which we seek
to bound by means of the synchrotron radiation.  We use the modified
dispersion relation (\ref{higher}) for the Hamiltonian of a charged
spinless particle probe, which we place in an external magnetic field.  
It is convenient to parametrize the QG effects ${p^{2+\alpha}}/{{\cal
M^\alpha}}$ by means of the refractive index in vacuo ${\cal \eta}$:
\begin{equation}
{\cal \eta} = 1 - \left(\frac{E}{{\cal M}}\right)^\alpha~, \qquad {\bf \alpha~\ge~1~}.
\label{etaalpha}
\end{equation} 
At ultra-relativistic energies, one may replace $p^\alpha$ by $E^\alpha$ 
in the dispersion relation, and thus use the following for the
Hamiltonian:
\begin{equation} 
E^2 = p^2 + m_0^2 -p^2\left(\frac{E}{{\cal M}}\right)^\alpha.
\label{mdr2}
\end{equation}
Notice that, in writing the dispersion relation in the form (\ref{mdr2}),
we have ignored terms proportional to the mass $m_0$ in the QG
modifications, expecting them to be subleading, which is a self-consistent
approximation at ultra-relativistic electron energies, as we shall see.
Moreover, for the range of energies relevant to our problem, we can also
ignore terms of order $\left(E/{\cal M}\right)^{2\alpha}$ or higher in the
following.

Before analyzing the consequences of (\ref{mdr2}), we first make some
important remarks about the Liouville foam model of~\cite{emn}, as opposed
to generic models of phenomenological modified dispersion relations in
flat space-times, such were considered in~\cite{jlm}. One important
difference of the Liouville model of foam is that the modifications in
the dispersion relation of a string probe owe their existence in the
appearance of a non-trivial induced target-space metric:
\begin{equation}  
G_{00}=-1~; \quad G_{ij}=\delta_{ij}~; \quad G_{0i} = g_s\Delta p_i /M_s\sim 
\frac{1}{2}g_sp_i/M_s 
\label{metricrec}
\end{equation}
where $M_s/g_s = {\cal M}$ is the effective QG scale, $M_s$ is the string
scale, $g_s$ a string coupling, and $\Delta p_i$ is the momentum transfer
during the scattering of the string with the D-particle defect~\cite{emn}.
A generic dispersion relation for a particle with mass $m_0$ in the metric
background (\ref{metricrec}) is:
\begin{equation} 
p_\mu p_\nu G^{\mu\nu} = -m_0^2~, 
\label{propcurve}
\end{equation} 
which, as can readily be seen, implies~\cite{emn} a modified 
dispersion relation of {\it subluminal} form (\ref{mdr2}), 
in the case where the direction of recoil of the D-particle defect
is opposite to that of the incident particle, in the frame where the 
(massive) defect was initially at rest, 
\begin{equation}
{\vec p} \cdot u^i \sim \frac{{\vec p} \cdot {\vec p}}{2{\cal M}} < 0~. 
\label{opposite}
\end{equation}
We recall that the requirement of subluminality emerges from specific
properties of the underlying string theory, namely the Born-Infeld form of
the effective action describing the dynamics of the excitations of the
recoiling D-particle~\cite{emn}.

Physically, this opposite recoil corresponds to capture of the particle
probe with subsequent decay of the defect, and emission of a particle with
modified momentum~\footnote{Ref.~\cite{amelino} considered synchrotron
radiation in terms of an interaction $e^- + \gamma_H \to e^- + \gamma$,
where $\gamma_H$ denotes a soft photon representing the magnetic field. In
our approach, we consider the quantum theory of synchrotron
radiation~\cite{Sokolov:nk}, in which an electron in a classical magnetic
field emits a quantum of synchrotron radiation by changing the state of
its wave function. This emission depends on a three-particle vertex that
has been discussed in the context of our model of quantum gravity
in~\cite{emngzk}, and could in principle violate energy conservation.
However, such effects are important only when all three of the quantum
particles have large momenta, which is not the case here, since the
emitted photon is relatively soft.}. It is this feature that may lead to
violations of the equivalence principle, if electrons and photons have
different probabilities for this process. This indeed happens in simple
models, as we discuss later.

We also observe from (\ref{propcurve}) that the inverse of the metric
(\ref{metricrec}) results in a higher-order `renormalization' of the
rest-mass term $m_0^2$ in the dispersion relation (\ref{mdr2}): $m_0^2 \to
m_0^2 ( 1 + |{\vec u}|^2)$. We ignore such higher-order effects here,
which is a self-consistent approximation for the range of energies
relevant to the problem.

In what follows, we keep our analysis as generic as possible, 
so as to minimize the 
dependence on specific models of foam. To this end, we use
the modified dispersion relation (\ref{mdr2})  to define the momentum
operator in a way that takes properly into account the QG
modification~\footnote{In the Liouville foam model~\cite{emn}, this
definition could be thought of as the gravitational covariant derivative
associated with the non-Minkowski induced metric (\ref{metricrec}).}:  $p
\to {\tilde p} \equiv p \eta$, which becomes ${\tilde p} = p\left(1 -
\frac{E}{{\cal M}}\right)$ in the linear case, where the minus sign indicates
subluminal propagation. This definition implies that we define an
`effective' momentum squared as the part one has to subtract from the
square of the energy in order to obtain the rest mass
squared~\footnote{This definition resembles that in general relativity,
where one defines the effective potential in a curved Schwarzschild
space-time as the part one has to subtract from the square of the energy
in order to obtain the square of the radial kinetic energy term.}.

It is this ${\tilde p}$ that we shall couple to the external
electromagnetic field, ${\vec A}$, ensuring gauge invariance by minimal
substitution, which implies the electromagnetic covariantization
\begin{equation} 
{\bf {\tilde p}} \to {\bf {\tilde p}} + \frac{e_0}{c}{\bf {\vec A}}.
\label{cov2}
\end{equation} 
As we now show in detail, in the case of QG-induced modifications of the
electron and photon dispersion relations, the maximal synchrotron
frequency (\ref{sync})  gets modified, because of the modification of the
orbital frequency {\it as well as} the velocity factor $\beta$ of the
electron:
\begin{equation} 
\omega_1 = \frac{\omega^{QG}_0}{1 - \beta_{QG}},
\label{syncqg}
\end{equation}
where we now concentrate on the longitudinal 
motion. 

The modification of the orbital frequency is connected with QG
effects on the radius of the orbit, due to the back-reaction of the foamy
QG medium on the propagation of the electron. Such effects are purely
quantum in nature, and therefore require a quantum treatment of the motion
of an electron of charge $e=-e_0$ in a magnetic field, which we shall now
present. For our purposes of estimating the change in the radius of the
orbit as a result of the QG modifications of the dispersion relation, we
do not take into account the spin of the electron, but consider instead
the simpler problem of a spinless particle in a magnetic field. We justify
the validity of this simplification later in the article.

We consider a magnetic field along the $z$ axis, which implies a vector
potential $A_x = -\frac{1}{2}yH~, \quad A_y = \frac{1}{2}xH~, \quad A_z =
0$, and use the QG-induced dispersion relation (\ref{mdr2}) for the
Hamiltonian operator of the quantum mechanical problem that describes the
coupling of a spinless particle with this potential.  The associated
Klein-Gordon equation \cite{Sokolov:nk} can then be written in the following form
(here we state explicitly the dependences on the speed of light in vacuo 
$c$):
\begin{equation}
\{ E^2 - c^2(-{\vec p}{\cal \eta} + \frac{e_0}{c}{\vec A})^2 - m_0^2c^4\}\Psi 
({\vec r}, t) =0~, \quad {\vec p} \equiv -i\hbar {\vec \nabla},
\label{mkg}
\end{equation}
where ${\cal \eta}$ is defined in (\ref{etaalpha}).  Proceeding, as usual,
to the stationary case via \\ $\psi({\vec r}, t) = {\rm exp}(-i\frac{E}{\hbar}
t)\psi ({\vec r})$, we have:
\begin{equation} 
\left(k^2 - k_0^2 + {\cal \eta}^2 \nabla ^2 - 
{\cal \eta}\frac{2ie_0}{c\hbar}{\vec A}\cdot {\vec \nabla} - 
\frac{e_0^2}{c^2\hbar^2}A^2\right)\psi({\vec r}) = 0,
\label{stationary}
\end{equation}
where $k={E}/{c\hbar}$ and $k_0={m_0c}/{\hbar}$. In what follows,
we consider only positive values of the energy, i.e., $k >0$, in which
case we may rewrite (\ref{stationary}) in terms of cylindrical coordinates
$(r, \varphi, z$), as follows:
\begin{equation}
\left( k^2 - k_0^2 + {\cal \eta}^2 \nabla_r ^2 +  
{\cal \eta}^2 \frac{\partial^2}{\partial z^2} + \frac{{\cal 
\eta}^2}{r^2}\frac{\partial ^2}{\partial \varphi^2} - 2i{\cal \eta}\gamma 
\frac{\partial}{\partial \varphi} - \gamma^2{\cal \eta}^2\right)\psi 
(r,\varphi, z) = 0,
\label{kgcylin}
\end{equation} 
where $\gamma \equiv {e_0 H}/{2c\hbar}$. 

Following~\cite{jlm}, we now assume that the $z$ component of momentum and
the angular momentum are conserved as a consequence of rotational
symmetry, which is maintained in the analysis of \cite{jlm}~\footnote{In
certain models of anisotropic QG foam, angular momentum is not necessarily
conserved. In such a case, our analysis here should be modified.}. Changing
variables to $\varphi_1 = \varphi/{\cal \eta}$ and ${\tilde z} = z/{\cal
\eta}$, we make the ansatz:
\begin{equation} 
\psi = \frac{e^{i\ell \varphi_1} e^{ik_3 {\tilde z}}}{(2\pi)^{1/2} L^{1/2}}
f(r),
\label{radial1}
\end{equation}
where $k_3 = 2\pi n_3/L$, and $\ell$ and $n_3$ are the azimuthal and
vertical quantum numbers (corresponding to the appropriate components of
the orbital angular momentum), which may take on positive and negative
values, including zero. The radial part of the wave-function satisfies:
\begin{equation} 
\{ \rho \frac{d^2}{d\rho^2} + \frac{d}{d\rho} + \lambda - \frac{\ell}{2}
-\frac{{\cal \eta}^2\rho}{4} - \frac{\ell ^2}{4\rho{\cal \eta}^2}\}f = 0,
\label{radial}
\end{equation}
where $\rho \equiv {\gamma r^2}/{{\cal \eta}^2}$, and $\lambda =
({k^2 - k_0^2 - k_3^2})/{4\gamma}$.

The asymptotic behaviour as $\rho \to \infty$ of the radial function $f(\rho)$
can be chosen such that 
\begin{equation} 
f_\infty = f(\infty) \sim e^{-\frac{\rho}{2}}~, \quad 
f_0 = f(0) \sim p^{\ell/2}~. 
\end{equation} 
This allows us to write 
\begin{equation}
f = f_\infty f_0 u(\rho) = e^{-(\rho/2)}~p^{\ell/2}~u(\rho),
\label{ueq}
\end{equation}
where $u(\rho )$ satisfies the equation
\beq 
\rho u'' + (\ell + 1 -\rho)u' + \left(\lambda - \ell -\frac{1}{2} + 
\frac{\rho}{4}
(1 - {\cal \eta}^2) + \frac{\ell^2}{4\rho}(1 - \frac{1}{{\cal 
\eta}^2})\right)u =0.
\label{radialu}
\eeq
We now write this equation in a form 
that can be compared with the confluent hypergeometric equation
\beq
\label{confhyp}
\rho u'' + (\kappa -\rho)u' - \left(a + a_1\rho + \frac{a_2}{\rho}\right)u 
=0,
\eeq
where we set $\kappa=\ell +1$, $a=-\lambda +\ell +\frac{\ell}{2}$,
$a_1=-\frac{1-\eta^2}{4}$ and $a_2=\frac{\ell^2}{4}(1-\frac{1}{\eta^2})$.  
Making the substitution $u=g\rho^{-\kappa/2}e^{\rho /2} $  and 
changing the variable $\rho 
=\frac{\tilde\rho}{\sqrt{4a_1+1}}$ in (\ref{confhyp}), we obtain 
\beq
\label{whiteq}
g''+\left[ -\frac{1}{4}+\frac{\frac{\kappa}{2}-a}{\tilde\rho\sqrt{4a_1+1}}+
\frac{\frac{\kappa}{2}-\frac{\kappa^2}{4}-a_2}{\tilde\rho^2} \right]g=0.
\eeq
Substituting ${\bar 
k}=({\frac{\kappa}{2}-a})/{\sqrt{4a_1+1}}$, 
$m=\sqrt{\frac{1}{4}-\frac{\kappa}{2}+
\frac{\kappa^2}{4}+a^2}$, we
reduce (\ref{whiteq}) to the form of a standard Whittaker equation, 
\beq
\label{standardw}
g''+\left[ -\frac{1}{4}+\frac{\bar k}{\tilde\rho}+\frac{\frac{1}{4}-m^2}{\tilde\rho^2} 
\right]g=0.
\eeq
The solution for positive energies is
\beq
\label{solution}
g=\tilde\rho^{m+1/2}e^{-\tilde\rho /2}{_1}F_1(m+\frac{1}{2}-\bar k; 2m+1; \tilde\rho ),
\eeq
where ${_1}F_1(A,B,\tilde\rho )$ is the confluent hypergeometric function, which
grows exponentially as $\rho \to \infty$.  Therefore, the asymptotic
behavior of the solution of (\ref{radialu}) for large $\rho$ takes the
form
\beq 
\label{asympt}
u\simeq e^{\rho}\rho^{1/2-\lambda}\frac{\Gamma (B)}{\Gamma (A)},
\eeq
where $\Gamma (x)$ is the standard Gamma function. This implies that, in
order to respect the finiteness of the probability of the wave-function as
$\rho \to \infty$, one has to impose the vanishing of the factor
$1/\Gamma(m+\frac{1}{2} - {\bar k})$, i.e., the selection rule:
\begin{equation}\label{selection}
m+\frac{1}{2}-\bar k = -s~, \quad s = 0,1,2, \dots 
\end{equation}
where $s$ is the well-known radial quantum number. Substituting the value 
of $\lambda$ given above, c.f., (\ref{radial}), we obtain: 
\begin{equation}
\label{spectrumqg}
\frac{k^2 - k_0^2 - k_3^2}{4\gamma} = 
\frac{1}{2}\ell\left(\frac{(2{\cal \eta}^2 - 1)^{1/2}}{{\cal \eta}} + 
1\right) + s{\cal \eta} + \frac{{\cal \eta}}{2}.
\end{equation}
{}From this expression, we see that our analysis is valid only in the regime
of energies for which ${\cal \eta}^2 > 1/2$, which implies
\begin{equation} 
(E/{\cal M})^\alpha < 1 - \frac{1}{\sqrt{2}} \simeq 0.293~. 
\label{limit}
\end{equation} 
This is not unreasonable, given that the concept of the effective field
theory breaks down at the scale ${\cal M}$. Within this range of energies, one
can justify self-consistently the approximation of ignoring terms of order
$\left(E/{\cal M}\right)^{2\alpha}$ or higher.

We now compare the QG-modified spectrum (\ref{spectrumqg}) with that
appearing in the standard Lorentz-invariant theory, namely: 
\beq
\label{spectrum} \frac{k^2 - k_0^2 - k_3^2}{4\gamma} = \ell +s-
\frac{1}{2}=\tilde n+ \frac{1}{2}, 
\eeq 
where ${\tilde n} =0,1,2, \dots$ is the principal quantum number.  In the
relativistic case of interest to us, both the spectra (\ref{spectrum}) and
(\ref{spectrumqg}) become quasi-continuous, since the quantum number
$\tilde n$ becomes very large.  QG corrections simply introduce a small
rescaling (by a factor of order one)  of the eigenvalues of the energy
states which define the quantized orbiting radii.

Assuming that the motion is in the plane of the orbit ($k_3 =0$), one can
then determine the radius $R$ of the orbit just by equating the energy
eigenvalues as given by (\ref{spectrumqg}) with those obtained in the case
of a particle in a magnetic field:
\begin{equation} 
c\hbar \left[ 4\gamma \left( \frac{1}{2}\ell\left(\frac{(2{\cal \eta}^2 - 
1)^{1/2}}{{\cal \eta}} + 1\right) + s{\cal \eta} + 
\frac{{\cal \eta}}{2}\right)\right]^{1/2} = e_0HR_{QG},
\label{approx}
\end{equation}
from which we obtain 
\begin{equation} 
R_{QG} = \left(\frac{\frac{1}{2}\ell\left(\frac{(2{\cal \eta}^2 - 
1)^{1/2}}{{\cal \eta}} + 1\right) + s{\cal \eta} + 
\frac{{\cal \eta}}{2}}{\gamma}\right)^{1/2}.
\label{radiusqg}
\end{equation}
for the QG-modified radius $R_{QG}$.

We are interested in macroscopic values of the orbit radius, which, in the
Crab Nebula case discussed in~\cite{jlm}, lie in the range $\simeq
10^{10}$cm. Comparing this with the other scales in the problem, we see
that we are in a situation where $\tilde n >> 1$. Thus, the equation
(\ref{approx})  can be approximated by
\begin{equation} 
\frac{R_{QG}}{R_0} \approx  \frac{1}{\sqrt{2}}\left(1 + \sqrt{2 -
\frac{1}{{\cal \eta}^2}}\right)^{1/2}~,
\label{modfinal}
\end{equation} 
where 
\begin{equation} 
R_0 =\left(\frac{{\tilde n} + \frac{1}{2}}{\gamma}\right)^{1/2}
\label{ro}
\end{equation} 
is the radius in the Lorentz-invariant theory. 

Some comments about this expression are now in order. First, we remark on
the functional dependence of $R_0$ on the principal quantum number
${\tilde n}$. In standard quantum mechanics, the mean square radius for a
particle of spin $\s$ in a magnetic field is given by: $ {\overline R_0} =
\sum_\xi \int \Psi^*_{{\tilde n},s,\xi} \Psi_{{\tilde n},s,\xi} r d^3r $,
where $\Psi_{{\tilde n}s}$ is the wave function, and the sum is over
possible spin states $\xi$. In the case of the Klein-Gordon equation,
$R_0$ is obtained by solving (\ref{radialu}) exactly in the limit ${\cal
\eta } =1$, with the result
\begin{equation} 
{\overline R_0^{KG}} = \left(\frac{{\tilde n}}{\gamma}\right)^{1/2}\left(1 + 
\frac{s + \frac{3}{2}}{4 {\tilde n}}\right).
\label{KG}
\end{equation} 
On the other hand, in the case of a particle with spin $s=1/2$, 
corresponding 
to the Dirac equation in an external field, one obtains \cite{Sokolov:nk}:
\begin{equation} 
{\overline R_0^{{\rm Dirac}}} = 
\left(\frac{{\tilde n}}{\gamma}\right)^{1/2}\left(1 + 
\frac{s + \frac{1}{2}}{4 {\tilde n}}\right).
\label{Dirac}
\end{equation} 
If one is interested in macroscopically large trajectories, as is the case
for~\cite{jlm} and ourselves, where ${\tilde n} \gg 1$, we see from
(\ref{KG}),(\ref{Dirac}) and (\ref{ro}) that all three formulae give
approximately the same result. In particular, the spin dependence of the
shape of the orbit is not significant. This justifies our initial
simplification of working with a spinless particle instead of a Dirac
fermion in order to estimate the QG modifications to the electron orbit.

This analysis demonstrates that QG modifies the critical
synchrotron frequency (\ref{syncqg}) for radiation along the forward
direction, in two ways. First by changing the relation of $\beta$ to
energy, due to the QG modification of the dispersion relation, but also by
modifying the orbit of the particle in the magnetic field, by slightly
shrinking the average radius (\ref{modfinal}). Thus, we may parametrize
the maximal frequency (\ref{sync})  as follows:
\begin{equation} 
\omega_c^{QG} = \frac{3}{2}\frac{e H}{m_0}\frac{1}{1 - \beta_{QG}^2}
\frac{R_0}{R_{QG}},
\label{qgparam}
\end{equation}
where the dependence on $R_{QG}$ is dictated by the assumption of angular
momentum conservation - the smaller the radius the higher the angular
frequency. It is important to notice that, once one accepts the dispersion
relation (\ref{mdr2}), the result (\ref{KG}) should be considered as {\it
exact} for the {\it entire} range of energies for which (\ref{limit}) is
valid, up to terms of higher order in $(E/{\cal M})^{\alpha}$.

\begin{figure} [tb]
\begin{center}
\epsfig{file=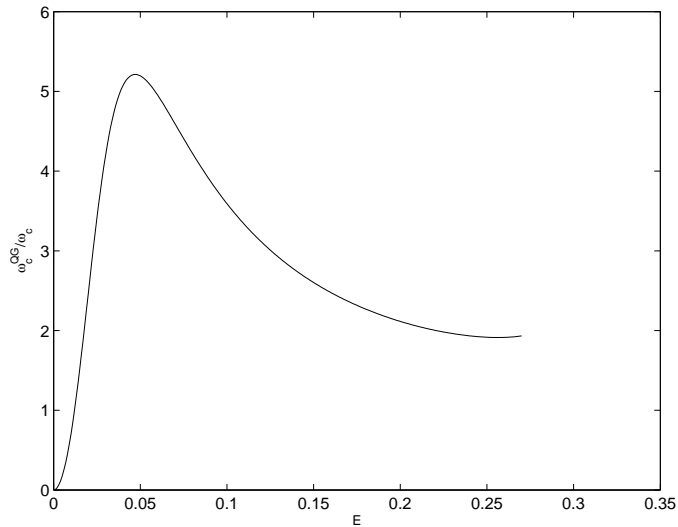,width=0.6\textwidth}
\end{center}
\caption{\it  
The scaled critical frequency for synchrotron radiation in quantum
gravity: $\omega_c^{QG}/\omega_c$ ($\omega_c=3eH/2m_0$) as a function of
the electron energy $E$ in units of ${\cal M}$, as obtained from 
(\ref{modqg}) for $\alpha=1$. The synchrotron frequency is a bounded
function of $E$, as in~\cite{jlm}, a feature which is not affected by the
QG modifications of the orbit.}
\label{fig:distribution}
\end{figure}

Using the modified dispersion relation, then, identifying $\beta^{QG}$
with the group velocity, and keeping only terms linear in 
$(E/{\cal M})^\alpha )$, we obtain the following result for the group
velocity:
\begin{eqnarray}  
&~&\beta_{QG} 
\equiv \frac{\partial E}{\partial p} = \frac{p}{E}
\left(1 - \left(\frac{\alpha}{2} + 1\right)
\left(\frac{E}{{\cal M}}\right)^\alpha\right)~, \nonumber \\ 
&~& 1 - \beta_{QG}^2 \simeq \frac{m_0^2}{E^2} + 
(\alpha + 1)\left(\frac{E}{{\cal M}}\right)^\alpha~.
\label{bfactor}
\end{eqnarray}
On account of (\ref{modfinal}), then, this yields for the QG modification
to the critical synchrotron radiation frequency:
\begin{equation}
\label{synchrohigher}
\omega_c^{QG} = \frac{3}{\sqrt{2}}\frac{e H}{m_0}\frac{1}{
(1 + \sqrt{2 -1/{\cal \eta}^2})^{1/2}
\left(\frac{m_0^2}{E^2} + (\alpha + 1)\left(\frac{E}{{\cal M}}\right)^\alpha
\right)}.
\label{modqg} 
\end{equation} 
This function is plotted schematically (for $\alpha =1$) 
in Fig.~\ref{fig:distribution}. 

We observe that, since the effects on the radius are small for all values
of $E/{\cal M} <0.3$, the discussion of~\cite{jlm} reviewed in the
previous Section remains qualitatively correct.  In particular, there is a
global maximum within the range of energies allowed by (\ref{limit}),
namely $\omega_{c,max}^{QG}$, which should be higher than the observed
synchrotron energy of $0.5$ GeV.  
The maximum occurs for 
electron energies $E_{max} = \left({2m_0^2 {\cal M}^\alpha}/{\alpha 
(\alpha + 1)}\right)^{1/(2+\alpha)}$.
From this, one may obtain bounds on
$\alpha$, if one sets ${\cal M} = M_P \sim 10^{19}$ GeV, or,
alternatively, obtain bounds for $|\xi_e|$ if one sets $\alpha$ to a fixed
value. 
 It can be easily seen 
that the general formula (up to terms of order one) for the upper bound on
$|\xi_e|$  is:
\beq
|\xi_e| < 
\left(\frac{3eH}{m_0}\right)^{\frac{\alpha + 2}{2\alpha }}\left(\frac{M_P}{m_0}\right) 
\left(\frac{2}{\alpha (\alpha + 
1)}\right)^{1/\alpha}\left(\frac{\alpha}{\alpha + 2}\right)^{(\alpha + 
2)/2\alpha}~. \label{generalalpha}
\eeq
If the resulting bound on $|\xi_e|$ is less than unity, the sensitivity
exceeds the Planck scale for the given value of $\alpha$. For the linear
case $\alpha =1$, the existence of the maximum implies that the discussion
of~\cite{jlm} remains intact and one arrives at the bounds of $\xi_e$
given in (\ref{eq:synchcon}),(\ref{eq:numba2}).

It is easy to see that 
for the average magnetic 
field of Crab Nebula~\footnote{It should be emphasized that the 
estimate of the end-point energy of the Crab synchrotron spectrum 
and of the magnetic field used above are indirect values based on 
the predictions of the Synchrotron Self-Compton (SSC) model 
of very-high-energy emission from Crab Nebula~\cite{reports}. We use
a choice of parameters which gives good agreement 
between the experimental data on high-energy emission and the 
predictions of the SSC model \cite{reports,hilas}.} 
$H_{cons}=260\mu{\rm G}$ \cite{magfield} one obtains $|\xi_e| 
< 6.8\cdot 10^{21}
\left( 8.8\cdot 10^{-21}\right)^{\frac{\alpha +2}{2\alpha}}$, 
which implies that $\alpha \le 1.72$ is excluded. 
If the lower value for $H_{ncons}=160\mu {\rm G}$ \cite{reports} is used instead, 
then $|\xi_e| \le 1$ for $\alpha \le 1.74$; we also observe that for 
$\alpha = 2$ 
$|\xi_e| < {\cal O}(30-60)$ for the range of the 
magnetic field considered above. These  
imply already a sensitivity 
to quadratic QG corrections.  
Therefore, 

\begin{equation} 
\alpha \ge \alpha_c~: \qquad 1.72 < \alpha_c < 1.74
\label{alphabounds}
\end{equation}
is the phenomenologically allowed range of $\alpha$, where the lower
(upper) limit corresponds to $H_{cons}$ ($H_{ncons}$).

We should note that there are still some ambiguities regarding the
magnitude of the maximal energy of synchrotron radiation before transition
to inverse Compton emission, which could range from 30 MeV~\footnote{
If the end-point of the Crab synchrotron spectrum is as low as 30~MeV, 
then
the upper and lower limits for $\alpha_c$ in \form{alphabounds} become
$1.56$ and $1.58$ respectively.} to 0.5 GeV. This uncertainty should be
resolved by future gamma-ray detectors such as GLAST~\cite{stecker}, which
could provide a better determination of the unpulsed gamma-ray spectrum in
the energy range above $30$ MeV, and thus a more precise determination of
the maximum electron energy in the Crab Nebula.

Measurements of high-energy emissions from the Vela pulsar~\cite{vela}
indicate the operation of a mechanism for the production of
very-high-energy gamma rays similar to that in Crab Nebula. This leads to
an estimate of the magnetic field of order $H_{Vela}\sim 3\mu{\rm G}$.
When used in (\ref{generalalpha}), this yields $\alpha_c=2.04$, under the
assumption that the first synchrotron hump is at the same location as in
the Crab Nebula, namely 0.5~GeV. If, one the other hand, the hump is at
30~MeV, then $\alpha_c=1.86$. Therefore, improvements in measurements of
synchrotron emission from the Vela Nebula (or other similar sources),
which may be achieved in the not-too-distant future, could bring the
sensitivity of such experiments closer to testing quadratic modifications
of the electron dispersion relation.

This will be decisive for probing models with quadratic
suppression of QG effects in the dispersion relation for {\it electrons}.
However, we remind the reader that synchrotron radiation provides no
significant bounds for QG modifications to the dispersion relation for
{\it photons}~\cite{jlm}.

\section{Violations of the Equivalence Principle in the Liouville Model for 
Space-Time Foam}

 \begin{figure} [tb]
\begin{center}
\epsfig{file=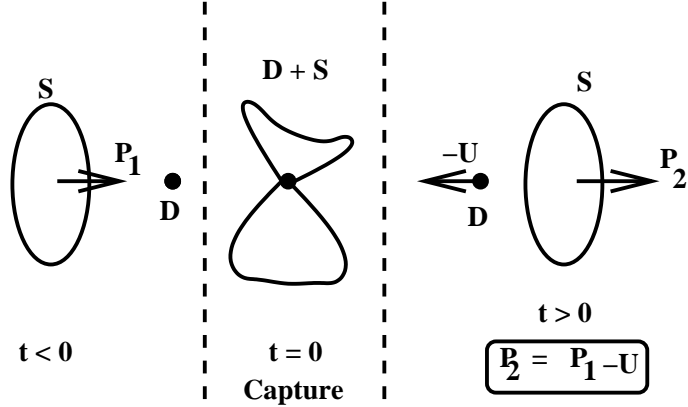,width=0.6\textwidth}
\end{center}
\caption{{\it  In the model of Liouville foam of \cite{emn}, 
only string particles (S) neutral under the (unbroken) 
standard model group can 
be captured by the D-particle defects (D) in space time. This results
in modified (subluminal) dispersion relations for S, as a consequence of 
the recoil of the defect D in a direction opposite to the 
incident beam, after the capture stage.}}
\label{fig:recoil}
\end{figure}

As already mentioned, the synchrotron radiation constraint~\cite{jlm} on
the dispersion relation for {\it electrons}, refined in the previous
Section, does not exclude the possibility of a larger QG modification of
the dispersion relation for {\it photons}. Of course, this would require
the violation of the equivalence principle for energetic particles.  
Remarkably, this is exactly what is predicted by our Liouville
string/D-particle model for space-time foam, according to which {\it only
gauge bosons} such as photons might have QG-modified dispersion relations,
and {\it not charged matter particles} such as electrons.

This difference may be traced to a cornerstone of D-brane physics, namely
that excitations which are {\it charged} under the gauge group are
represented by open strings with their ends attached to the
D-brane~\cite{polchinski}, and only {\it neutral} excitations are allowed
to propagate in the bulk space transverse to the brane.

Our D-particle model of QG foam is based on point-like defects in
space-time, that are nothing but zero-space-dimensional D-branes, embedded
in a four- (or higher-) dimensional bulk Minkowski space-time.  These
affect the propagation of closed-string states via a recoil
process~\cite{emn} (see Fig.~\ref{fig:recoil}), that is associated with
back-reaction effects in target space, which lead to the QG-modified
effective metric (\ref{metricrec}). In this picture, a closed-string state
propagating in the bulk can be captured by the defect and split into two
open-string excitations with their ends attached to the defect, which
later recombine to become a closed-string state. According to the 
above-mentioned property of D-branes, the open-string excitations
of the D-particle defect result in massless U(1)
excitations for a single defect, or massless gauge bosons 
in the adjoint representation of U(N) for a group of coincident 
recoiling D-particles~\cite{bound,szabo}.

In the original formulation of this D-particle model of QG
foam~\cite{emn}, the D-particles have no `hair', i.e., they have only
vacuum quantum numbers. Therefore, they can absorb closed-string states
only if they also carry no conserved charges. This would exclude any QG
medium effects on particles with electromagnetic or colour charge,
certainly including {\it electrons} and {\it quarks}. On the other hand, a
QG effect would be expected for the {\it photon}, which carries no
conserved charge. Because the electron has no interaction with the QG 
vacuum medium in this approach, it emits no {\v C}erenkov radiation, 
despite traveling faster than photons, thus avoiding the vacuum {\v 
C}erenkov radiation 
constraint considered by~\cite{Seth}.

Neutrinos are mainly doublets of the SU(2) subgroup of the Standard Model.
However, in generic seesaw models of neutrino masses, the light neutrinos
mix with ${\cal O}(m_W/M)$ components of gauge-singlet states, perhaps
opening the window for QG effects, depending on details of the model.
However, any such effects would be suppressed by at least a similar
factor. In the case of gluons, which carry non-Abelian charges, it is
possible that ensembles of D-particle defects would have a QG effect on
their dispersion relations, which would in general be unrelated to that on
the photon.

However, such a relation might appear if the U(1) of electromagnetism is
actually embedded in some simple GUT group such as SU(5)~\cite{GG}, so 
that the
photon becomes some combination of non-Abelian gauge fields, in which case 
QG
effects on its dispersion relation might be related to those of the
gluons.  On the other hand, there exist non-simple GUT models, such as
flipped SU(5) $\times$ U(1)~\cite{F5}, in which the photon contains a 
significant
component from outside the non-Abelian group factor, in which case the 
photon's
dispersion relation will in general differ from those of the gluons.

Even if the dispersion relations of individual gluons are modified by QG
effects, these may be suppressed for colour-singlet hadrons such as
protons. A QG modification of the proton's dispersion relation has been
suggested as a way of explaining the possible existence of UHECRs beyond 
the GZK cut-off~\cite{piran}, where
energy non-conservation in high-energy reactions might 
also become significant~\cite{ng,emngzk}. As
observed in~\cite{emngzk}, in the case of a linear QG modification of the
proton's dispersion relation similar to (2), the magnitude of the
analogous parameter $\xi_p$ would need to be of order $10^{-16}-10^{-17}$.
Such a suppression is (coincidentally?) ${\cal O}(\Lambda_{QCD} / M_P)$,
and such a suppression cannot be excluded at present, though a lot more
work is required before such a conclusion could be reached in the
framework of the D-particle model of~\cite{emn}. However, the above
discussion makes it clear that the details of the dynamics of the
interactions between particles and space-time foam are potentially 
important, and any
naive phenomenological assumption of the equivalence principle could be
misleading.

Before closing, we comment briefly on recent claims that modified
dispersion relations for photons would result in phase incoherence of
light from distant galaxies~\cite{incoh}. Based on this suggestion, these
authors proposed a stellar interferometry technique, and claimed to place
stringent bounds on the effective quantum-gravity scale for photons,
excluding linearly-modified dispersions. However, their arguments have
been criticized in~\cite{ng2}, on the basis that they overestimated the
induced incoherent effects by a large factor $(L/\lambda)^{\alpha}$, where
$L$ is the distance of the source, $\lambda$ is the photon wavelength, and
$\alpha$ is the parameter in the modified dispersion relation of the
photon probe. The correct amount of cumulative phase incoherence induced
by quantum gravity is~\cite{ng2}: $\delta \phi \sim
(L/\lambda)^{1-\alpha}$. Thus, for linear dispersion relations for photons
($\alpha =1$) such as those proposed in~\cite{nature,emn}, the technique
of~\cite{incoh} cannot be used.  
Moreover, in the case of the D-particle recoil model for Liouville 
foam~\cite{emn}
the re-emission of the photon from the recoiling D-particle (c.f. 
Fig. \ref{fig:recoil}) is accompanied by a (random) phase in the 
photon's wave-function. This would destroy any cumulative phase 
incoherence, in contrast to the claims in \cite{incoh}. 
Thus, a linear modification of the photon
dispersion relation cannot be excluded by this argument.

Our analysis emphasizes the interest in probing independently the
dispersion relation of the photon. The study of the arrival times of
photons from gamma-ray bursts~\cite{gray} still appears to be the best
experimental probe of any possible refractive index for photons, and
should be pursued further in the future.

The mechanism for violating the equivalence principle discussed in this
article may not be the only route to consistency with the current
experimental constraints. However, it is certainly one promising way, in
the sense that it does not invoke exceedingly small parameters, and
appears naturally within a class of stringy models of quantum
gravity~\cite{emn}. In other approaches to quantum gravity, such as loop
canonical quantum gravity~\cite{smolin}, one may encounter models which
feature linear modifications in the dispersion relations for matter
excitations that are characterized by extra small parameters which can be
bound by experiments. Such a scenario has been discussed recently
in~\cite{dotti}, where the structure of canonical (quantum) commutators
between momenta and position operators in loop-gravity models is modified
by extra scaling terms, described by a set of coefficients that are
essentially free parameters of the model. Such extra terms lead to linear
modifications in dispersion relations, but with coefficients that are
proportional to these parameters. The latter can then be bounded by
requiring consistency of the models with the current experimental
situation. In our opinion, the disadvantage of such an approach lies in
the unnaturally small values one obtains for such parameters in this case,
which is to be contrasted with our case above, where the numerical
coefficient in front of the linear modification of the photon dispersion
relation is naturally of order one.  However, we cannot exclude the
possibility this, or some other way yet to be invented, may be the way
that Nature evades the current experimental constraints on quantum
gravity.

\section*{Acknowledgements}

We would like to thank H.~Hofer for his interest and support. The work 
of N.E.M. is partially supported by a visiting professorship at
the University of Valencia (Spain), Department of Theoretical 
Physics, and by the European Union (contract HPRN-CT-2000-00152).

\end{document}